\documentclass[onecolumn]{aastex63}

\shorttitle{20caa}
\shortauthors{Soraisam et al.}
\usepackage[caption=false]{subfig}

\begin{document}

\title{AT2020caa: A Type Ia Supernova with a Prior Outburst or a Statistical Fluke?}

\correspondingauthor{Monika Soraisam}
\email{soraisam@illinois.edu}

\author{Monika~Soraisam}
\altaffiliation{Illinois Survey Science Fellow}
\affiliation{National Center for Supercomputing Applications, University of Illinois at Urbana-Champaign, Urbana, IL 61801, USA}
\affiliation{Department of Astronomy, University of Illinois at Urbana-Champaign, Urbana, IL 61801, USA}

\author[0000-0001-6685-0479]{Thomas~Matheson}
\affiliation{NSF's National Optical-Infrared Astronomy Research Laboratory, 950 N Cherry Ave, Tucson, USA}
\author[0000-0003-1700-5740]{Chien-Hsiu~Lee}
\affiliation{NSF's National Optical-Infrared Astronomy Research Laboratory, 950 N Cherry Ave, Tucson, USA}

\begin{abstract}
%150 words limit
We recently discovered an extragalactic transient, AT2020caa,  using the community alert broker ANTARES. This transient apparently exhibited two outbursts in a time span of a year (between 2020 and 2021). Based on a decade-long historical light curve of the candidate host galaxy, we rule out an activity from the galaxy nucleus to explain these outbursts. The measured peak magnitudes (assuming the known spectroscopic redshift of the candidate host galaxy) put AT2020caa in the realm of thermonuclear supernovae (SNe) or luminous core-collapse SNe. A handful of the latter are known to show prior outbursts (POs), thought to be linked to mass loss in massive stars. Using Gemini/GMOS, we obtained a spectrum of the current outburst that shows it to be a Type Ia supernova (SNIa). We examine the nature of AT2020caa's PO and conclude that it is likely a separate SN within the same galaxy. 
\end{abstract}

\section*{ }
\begin{figure}
\centering 
\subfloat[]{
  \includegraphics[width=50mm]{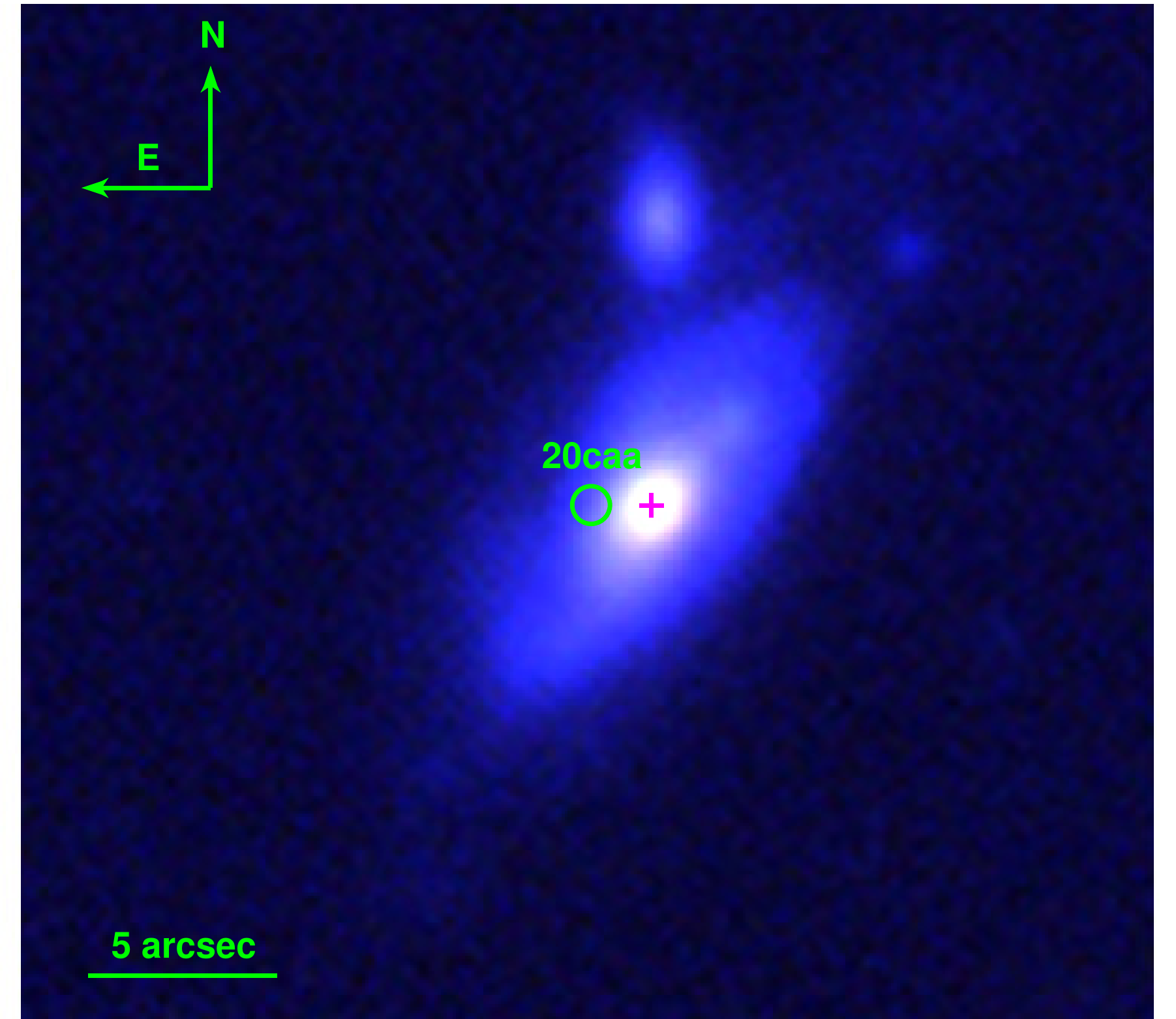}
}\quad
\subfloat[]{
  \includegraphics[width=140mm]{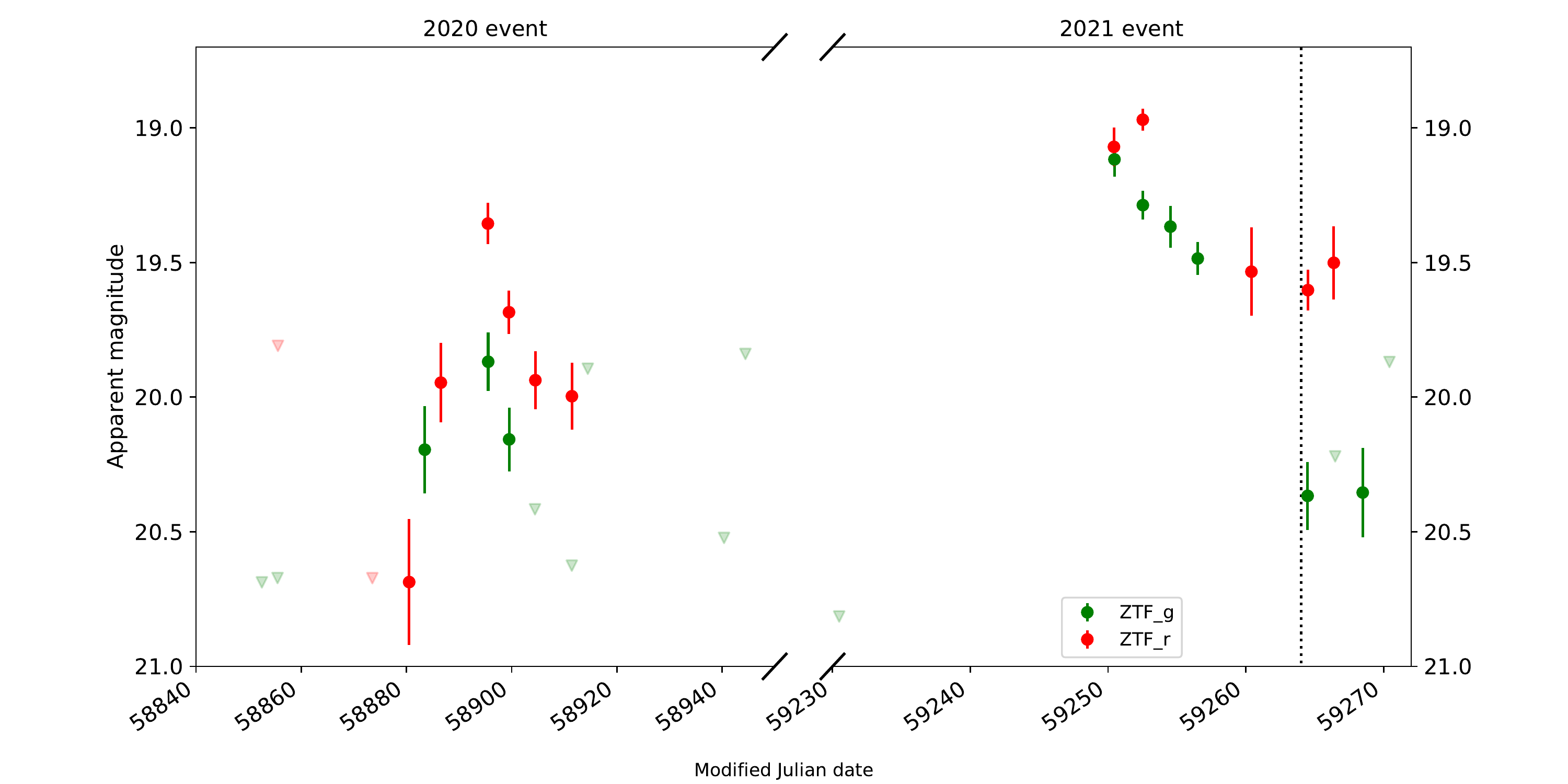}
}\quad
\subfloat[]{
  \includegraphics[width=140mm]{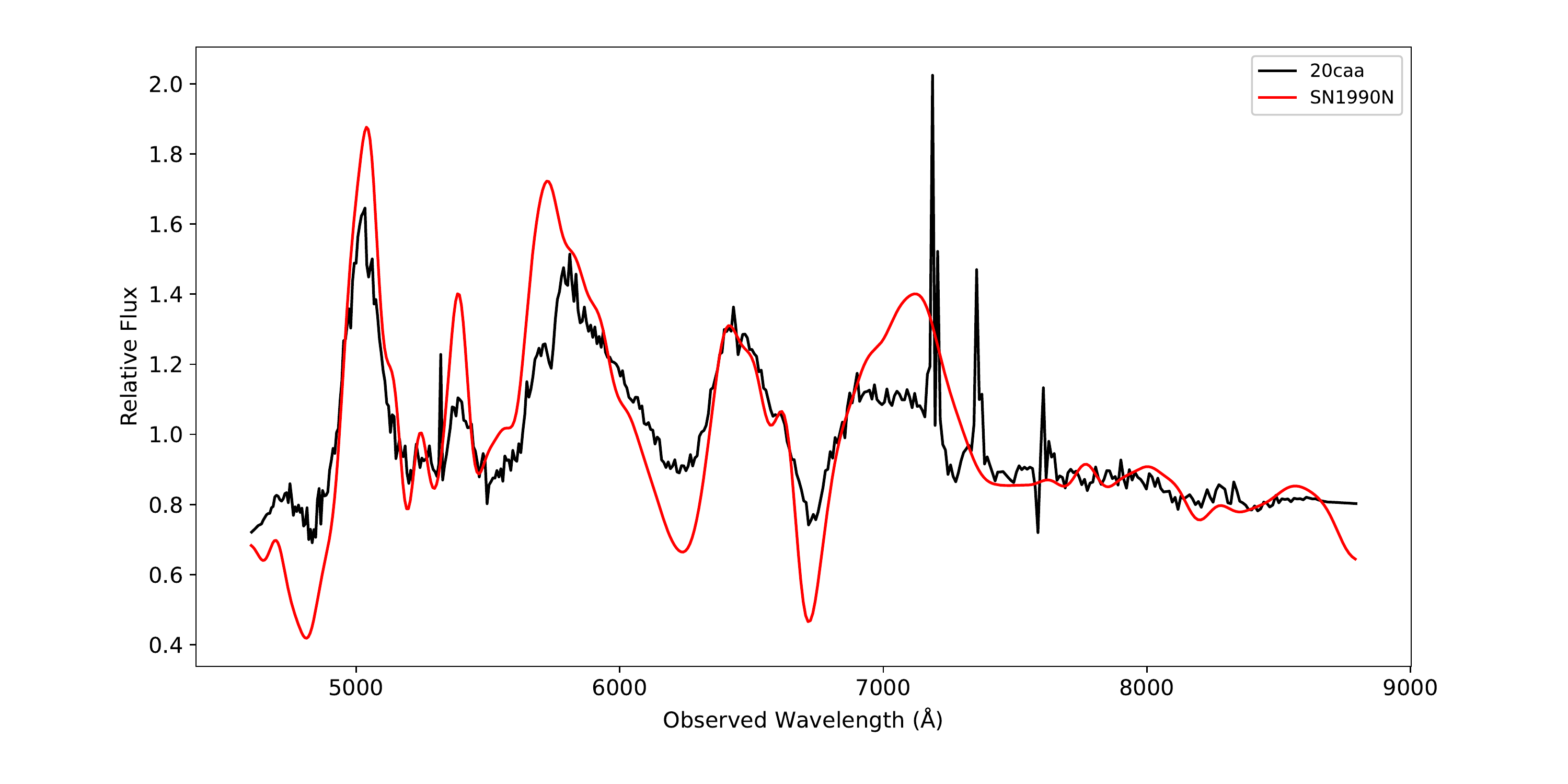}
}
\caption{{\it Panel (a)}: Color ($g,r,z$) cutout from the DESI Legacy Imaging Survey \citep{Dey-2019} showing the location of 20caa (green circle) with respect to the bright nucleus of the galaxy SDSSJ1419 (magenta +). Another galaxy, MGC 60395, can also be seen north of the former. 
{\it Panel (b)}: Light curve of 20caa based on forced-photometry of difference images from the ZTF survey. Triangles denote 5$\sigma$ upper limit while the dotted line marks the epoch when the spectrum was obtained for the 2021 event. Note the break in the horizontal axis; this is done to highlight the two outbursts. {\it Panel (c)}: Spectrum of 20caa obtained with Gemini-S GMOS, along with the best-fit SNIa spectrum (SN1990N; \citealt{Phillips-1992}).
}
\label{fig:offset}
\end{figure}

Modern wide-field, high-cadence optical surveys, such as the ongoing Zwicky Transient Facility (ZTF; \citealt{Bellm-2019}), regularly deliver thousands of supernovae (SNe) -- thermonuclear as well as core-collapse events -- per year.  
We discovered the transient event ZTF20aamibse{\footnote{\url{https://antares.noirlab.edu/loci/ANT2020l2ixy}}} (AT2020caa, hereafter 20caa) from the real-time alert stream of ZTF using the community alert-broker ANTARES \citep{Matheson-2021}. It is located at (J2000) RA 14h19m18.61s, Dec +00d03m27.86s, 1\farcs58 offset from the bright nucleus of a candidate host galaxy SDSS J141918.50+000327.8 (hereafter, SDSSJ1419) at a spectroscopic redshift of 0.097 (\citealt{sdss}; see Fig.~\ref{fig:offset} Panel a) and a distance of around 460~Mpc ($H_{0}=67.8, \Omega_{M}=0.307$; \citealt{planck-13}). 
In addition to the offset of 20caa from the nucleus, the decade-long light curve of this galaxy from the Catalina Sky Survey \citep{Drake-2009} does not show any variability, thus ruling out activity from an active galactic nucleus.
%The constant fit has reduced $\chi^{2}=0.88$, indicating that a more complex fit is not warranted. 

The ZTF light curve of 20caa based on forced photometry is shown in Fig.~\ref{fig:offset} (Panel b). It reveals a current outburst with an observed peak absolute magnitude $M_{r}=-19.4$~mag on MJD 59252 (2021 February 7 UT). This peak brightness puts 20caa in the realm of supernovae.
%(either thermonuclear or a very bright core-collapse SN with circumstellar interaction). 
Interestingly, it also showed a prior outburst (PO, hereafter) about a year before, just a few tenths of a mag fainter than the current outburst. The PO lasted (i.e., was above the detection threshold) for around a month. Note that this PO was also observed by the ATLAS survey \citep{Tonry-2018}{\footnote{Based on the ATLAS forced photometry service, the measurement reported by the survey to TNS %around MJD 58900 of ATLAS-$c$=19.05 (Fig.~\ref{fig:lc}) 
appears overestimated (possibly due to contamination by the host galaxy given the large pixel scale of 1\farcs86 for ATLAS). %There are three measurements within 1 hr of each other that yielded an average magnitude of ATLAS-$c$=19.4.
}}. Pre-explosion outbursts have been observed in a handful of core-collapse SNe \citep[e.g.,][]{Mauerhan-2013}. Their underlying mechanism is, however, not yet understood and is likely to be closely linked to ill-constrained processes in massive-star evolution such as mass-loss. Nevertheless, such POs are faint, typically peaking at $>-15$~mag (e.g., SN2010mc; \citealt{Ofek-2013}).

We obtained a spectrum for the 2021 event on 19 February (UT) with GMOS on Gemini South under a director's discretionary program (GS-2021A-DD-102; PI: Soraisam). The spectral signature is consistent with a Type Ia supernova (SNIa) at a phase of around 14 days past maximum, and a 
redshift of $0.095\pm0.005$ \citep[based on SNID,][]{Blondin-2007}, which agrees with that from the narrow emission lines in the spectrum and is consistent with the redshift of SDSSJ1419. We also triggered target-of-opportunity observations with {\it Swift}. In line with the SN~Ia interpretation, we did not detect the event in X-rays from two epochs of {\it Swift} observations on February 20 and 25 (UT).

\section*{Nature of the double event}
We explore three scenarios for the nature of the 2020 event of 20caa in light of the evidence we gathered for the current 2021 explosion. %We consider three possibilities as detailed below. 

\subsubsection*{Lensed SN?}
There is another galaxy near 20caa, MGC 60395 at a redshift of 0.0974 \citep{Liske-2015}, marginally larger than that of SDSSJ1419. %, separated by $7.8''$ from 20caa (seen at the top of the color cutout in Fig.~\ref{fig:offset}). 
The redshift of 20caa determined from its spectrum allows both SDSSJ1419 and MGC 60395 to be its host. Given that it is not possible to get effective lensing between two galaxies that are relatively close to one another as compared to their distances from the observer \citep[cf.~][]{Wambsganss-1998}, 
we rule out lensing for 20caa.

\subsubsection*{Prior outburst?}
The nature of SNIa progenitors is one of the biggest open questions in astronomy. Among the candidate progenitors, the single-degenerate scenario with an accreting white dwarf that produces recurrent novae is perhaps the only one that can be associated with historical outbursts of significant strength. Yet, the peak absolute magnitudes of novae reach just around -11~mag (see Fig.~B.1 in \citealt{Soraisam-2015}), too dim to account for the PO.
%For the limiting magnitude of ZTF, novae can therefore be detected only to a distance of around 16~Mpc.  

\subsubsection*{SN siblings?}
Multiple cases of sibling SNe, i.e., those sharing the same host galaxy, have been observed in the past \citep[e.g.,][]{Anderson-2013,
Scolnic-2020}. This is not surprising, given the typical overall SN rate of one per galaxy per century. However, for all of the ones with discernible extended host galaxies, the siblings are spatially well-separated. Note that a recent SN sibling from the DES survey, $\rm{DES16C3nd}_{0}$ and $\rm{DES16C3nd}_{1}$ \citep{Scolnic-2020}, occurred within 1\arcsec\ and temporally separated by 200 days. Nevertheless, at a relatively large redshift of 0.65, the angular extent of their host galaxy is comparable to 1\arcsec, which reconciles the spatial coincidence of $\rm{DES16C3nd}_{0}$ and $\rm{DES16C3nd}_{1}$.

Assuming the 2020 event of 20caa is a sibling SN, both of which occurred in the galaxy SDSSJ1419, we assess the likelihood of their spatial coincidence. The peak absolute magnitude $M_{r}$ of the 2020 event from ZTF is then $-19.04$ (corrected for Galactic extinction; \citealt{Schlafly-2011}). This peak brightness points to it being a SN~Ia (though some core-collapse, interacting SNe, SN1998s-like, were also found to reach such a brightness).

For ZTF observations, the typical FWHM of the image PSF is 2\arcsec, with a pixel scale of 1\farcs01 \citep{Bellm-2019}. The ANTARES broker conservatively uses a radius (ANT\_res) of 1\arcsec\ %(as compared to $1.5''$ used by the ZTF survey)
to aggregate alerts into a single locus, with the position of this locus assigned to the centroid of the group. 
For 20caa, there appears an offset of 1\farcs3 between the 2020 and 2021 events. This offset may be instrumental in nature, given the presence of the nearby bright nucleus that can influence difference imaging, or it may indeed be real. For the latter case, we determine below the probability of seeing a double SN in a galaxy with the two explosions occurring in such close angular proximity of each other.    

SDSSJ1419 is typed as a spiral Sab galaxy \citep[e.g.,][]{Huertas-2011}. We use the SNIa rate-size relation for galaxies of Sab morphology from \citet{Li-2011} and the stellar mass of SDSSJ1419, $\log(M/M_\odot)=10.38$ from \citet{Brinchmann-2004}, to compute its SNIa rate, which we obtain as $0.42^{+0.07}_{-0.07}$ per century. To be conservative, we also consider rates of core-collapse SNe based on \citet{Li-2011} ---  $0.38^{+0.12}_{-0.08}$ and $0.46^{+0.08}_{-0.07}$ per century for SNeIbc and SNeII, respectively --- and obtain a total SN rate of roughly $1.26$ per century.

Using the average Kron radius of SDSSJ1419 from PS1 DR2 \citep{Chambers-2016} to estimate the area of the whole galaxy and its total SN rate from above, we obtain the rate of SNe within a circular region of radius 1\arcsec\ (ANT\_res) as $1.8\times10^{-4}~{\rm yr}^{-1}$. For a given SN in this region, the probability of observing a second SN within a year by ZTF, $p(\Delta t<1~{\rm yr})=1.2\times10^{-4}$. Note that for the 3 day cadence of the public ZTF survey, its completeness for SN detection is largely determined by the visibility of the given galaxy, which we have factored into our computation. However, 20caa was selected from the output of a filtered stream (active since half a year ago) on ANTARES and therefore we need to account for the `look-elsewhere effect.' To this end, we compute a similar probability $p_{i}(\Delta t<1~{\rm yr})$ for each of the roughly $N=1000$ sources in this stream with SN-like light curve and discernible host-galaxy, assuming a typical SN rate of 1.5 per century for each galaxy. We then obtain the probability ($1-\prod_{i=1}^{i=N}(1-p_{i}(\Delta t<1~{\rm yr})$) of seeing a double event for one or more of the events in our filtered stream as 27\%.

In the absence of other possibilities (as discussed above), and given this appreciable probability for a chance encounter with such SN siblings, we consider the 2020 and 2021 events of 20caa as SN siblings. In this case, the spatial and temporal proximity of the siblings coupled with the fact that SDSSJ1419 is a star forming galaxy ($\log {\rm SFR}~(M_{\odot}~{\rm yr}^{-1}) = 0.6$; \citealt{Brinchmann-2004}) provide an opportunity to investigate their influence on the host galaxy local environment. Additionally, the 2020 SN appears likely an SNIa, in which case 20caa will be an addition to the growing sample of SNIa siblings that can be used to investigate the influence of host galaxy properties on SNIa-based distances. The fact that we could detect such spatially and temporally close SN siblings despite their rarity and having used the particular ANTARES alert stream that flagged it for only about half a year points to the power of large-scale time-domain surveys like ZTF when coupled with ANTARES' automated alert-filtering capabilities.

\acknowledgments
We thank Nathan Smith and Jennifer Andrews for attempting spectral observations with Binospec on MMT, as well as Adam Bolton for helpful discussions on gravitational lensing.

\bibliography{ref}{}

\begin{thebibliography}{}
\expandafter\ifx\csname natexlab\endcsname\relax\def\natexlab#1{#1}\fi
\providecommand{\url}[1]{\href{#1}{#1}}
\providecommand{\dodoi}[1]{doi:~\href{http://doi.org/#1}{\nolinkurl{#1}}}
\providecommand{\doeprint}[1]{\href{http://ascl.net/#1}{\nolinkurl{http://ascl.net/#1}}}
\providecommand{\doarXiv}[1]{\href{https://arxiv.org/abs/#1}{\nolinkurl{https://arxiv.org/abs/#1}}}

\bibitem[{{Ahn} {et~al.}(2012){Ahn}, {Alexandroff}, {Allende Prieto},
  {Anderson}, {Anderton}, {Andrews}, {Aubourg}, {Bailey}, {Balbinot}, {Barnes},
  {Bautista}, {Beers}, {Beifiori}, {Berlind}, {Bhardwaj}, {Bizyaev}, {Blake},
  {Blanton}, {Blomqvist}, {Bochanski}, {Bolton}, {Borde}, {Bovy}, {Brandt},
  {Brinkmann}, {Brown}, {Brownstein}, {Bundy}, {Busca}, {Carithers}, {Carnero},
  {Carr}, {Casetti-Dinescu}, {Chen}, {Chiappini}, {Comparat}, {Connolly},
  {Crepp}, {Cristiani}, {Croft}, {Cuesta}, {da Costa}, {Davenport}, {Dawson},
  {de Putter}, {De Lee}, {Delubac}, {Dhital}, {Ealet}, {Ebelke}, {Edmondson},
  {Eisenstein}, {Escoffier}, {Esposito}, {Evans}, {Fan}, {Femen{\'\i}a
  Castell{\'a}}, {Fern{\'a}ndez Alvar}, {Ferreira}, {Filiz Ak}, {Finley},
  {Fleming}, {Font-Ribera}, {Frinchaboy}, {Garc{\'\i}a-Hern{\'a}ndez},
  {Garc{\'\i}a P{\'e}rez}, {Ge}, {G{\'e}nova-Santos}, {Gillespie}, {Girardi},
  {Gonz{\'a}lez Hern{\'a}ndez}, {Grebel}, {Gunn}, {Guo}, {Haggard}, {Hamilton},
  {Harris}, {Hawley}, {Hearty}, {Ho}, {Hogg}, {Holtzman}, {Honscheid},
  {Huehnerhoff}, {Ivans}, {Ivezi{\'c}}, {Jacobson}, {Jiang}, {Johansson},
  {Johnson}, {Kauffmann}, {Kirkby}, {Kirkpatrick}, {Klaene}, {Knapp}, {Kneib},
  {Le Goff}, {Leauthaud}, {Lee}, {Lee}, {Long}, {Loomis}, {Lucatello},
  {Lundgren}, {Lupton}, {Ma}, {Ma}, {MacDonald}, {Mack}, {Mahadevan}, {Maia},
  {Majewski}, {Makler}, {Malanushenko}, {Malanushenko}, {Manchado},
  {Mandelbaum}, {Manera}, {Maraston}, {Margala}, {Martell}, {McBride},
  {McGreer}, {McMahon}, {M{\'e}nard}, {Meszaros}, {Miralda-Escud{\'e}},
  {Montero-Dorta}, {Montesano}, {Morrison}, {Muna}, {Munn}, {Murayama},
  {Myers}, {Neto}, {Nguyen}, {Nichol}, {Nidever}, {Noterdaeme}, {Nuza},
  {Ogando}, {Olmstead}, {Oravetz}, {Owen}, {Padmanabhan},
  {Palanque-Delabrouille}, {Pan}, {Parejko}, {Parihar}, {P{\^a}ris},
  {Pattarakijwanich}, {Pepper}, {Percival}, {P{\'e}rez-Fournon},
  {P{\'e}rez-R{\`a}fols}, {Petitjean}, {Pforr}, {Pieri}, {Pinsonneault}, {Porto
  de Mello}, {Prada}, {Price-Whelan}, {Raddick}, {Rebolo}, {Rich}, {Richards},
  {Robin}, {Rocha-Pinto}, {Rockosi}, {Roe}, {Ross}, {Ross}, {Rossi},
  {Rubi{\~n}o-Martin}, {Samushia}, {Sanchez Almeida}, {S{\'a}nchez},
  {Santiago}, {Sayres}, {Schlegel}, {Schlesinger}, {Schmidt}, {Schneider},
  {Schultheis}, {Schwope}, {Sc{\'o}ccola}, {Seljak}, {Sheldon}, {Shen}, {Shu},
  {Simmerer}, {Simmons}, {Skibba}, {Skrutskie}, {Slosar}, {Sobreira}, {Sobeck},
  {Stassun}, {Steele}, {Steinmetz}, {Strauss}, {Streblyanska}, {Suzuki},
  {Swanson}, {Tal}, {Thakar}, {Thomas}, {Thompson}, {Tinker}, {Tojeiro},
  {Tremonti}, {Vargas Maga{\~n}a}, {Verde}, {Viel}, {Vikas}, {Vogt}, {Wake},
  {Wang}, {Weaver}, {Weinberg}, {Weiner}, {West}, {White}, {Wilson},
  {Wisniewski}, {Wood-Vasey}, {Yanny}, {Y{\`e}che}, {York}, {Zamora},
  {Zasowski}, {Zehavi}, {Zhao}, {Zheng}, {Zhu}, \& {Zinn}}]{sdss}
{Ahn}, C.~P., {Alexandroff}, R., {Allende Prieto}, C., {et~al.} 2012, \apjs,
  203, 21, \dodoi{10.1088/0067-0049/203/2/21}

\bibitem[{{Anderson} \& {Soto}(2013)}]{Anderson-2013}
{Anderson}, J.~P., \& {Soto}, M. 2013, \aap, 550, A69,
  \dodoi{10.1051/0004-6361/201220600}

\bibitem[{{Bellm} {et~al.}(2019){Bellm}, {Kulkarni}, {Graham}, {Dekany},
  {Smith}, {Riddle}, {Masci}, {Helou}, {Prince}, {Adams}, {Barbarino},
  {Barlow}, {Bauer}, {Beck}, {Belicki}, {Biswas}, {Blagorodnova}, {Bodewits},
  {Bolin}, {Brinnel}, {Brooke}, {Bue}, {Bulla}, {Burruss}, {Cenko}, {Chang},
  {Connolly}, {Coughlin}, {Cromer}, {Cunningham}, {De}, {Delacroix}, {Desai},
  {Duev}, {Eadie}, {Farnham}, {Feeney}, {Feindt}, {Flynn}, {Franckowiak},
  {Frederick}, {Fremling}, {Gal-Yam}, {Gezari}, {Giomi}, {Goldstein},
  {Golkhou}, {Goobar}, {Groom}, {Hacopians}, {Hale}, {Henning}, {Ho}, {Hover},
  {Howell}, {Hung}, {Huppenkothen}, {Imel}, {Ip}, {Ivezi{\'c}}, {Jackson},
  {Jones}, {Juric}, {Kasliwal}, {Kaspi}, {Kaye}, {Kelley}, {Kowalski},
  {Kramer}, {Kupfer}, {Landry}, {Laher}, {Lee}, {Lin}, {Lin}, {Lunnan},
  {Giomi}, {Mahabal}, {Mao}, {Miller}, {Monkewitz}, {Murphy}, {Ngeow},
  {Nordin}, {Nugent}, {Ofek}, {Patterson}, {Penprase}, {Porter}, {Rauch},
  {Rebbapragada}, {Reiley}, {Rigault}, {Rodriguez}, {van Roestel}, {Rusholme},
  {van Santen}, {Schulze}, {Shupe}, {Singer}, {Soumagnac}, {Stein}, {Surace},
  {Sollerman}, {Szkody}, {Taddia}, {Terek}, {Van Sistine}, {van Velzen},
  {Vestrand}, {Walters}, {Ward}, {Ye}, {Yu}, {Yan}, \& {Zolkower}}]{Bellm-2019}
{Bellm}, E.~C., {Kulkarni}, S.~R., {Graham}, M.~J., {et~al.} 2019, \pasp, 131,
  018002, \dodoi{10.1088/1538-3873/aaecbe}

\bibitem[{{Blondin} \& {Tonry}(2007)}]{Blondin-2007}
{Blondin}, S., \& {Tonry}, J.~L. 2007, \apj, 666, 1024, \dodoi{10.1086/520494}

\bibitem[{{Brinchmann} {et~al.}(2004){Brinchmann}, {Charlot}, {White},
  {Tremonti}, {Kauffmann}, {Heckman}, \& {Brinkmann}}]{Brinchmann-2004}
{Brinchmann}, J., {Charlot}, S., {White}, S.~D.~M., {et~al.} 2004, \mnras, 351,
  1151, \dodoi{10.1111/j.1365-2966.2004.07881.x}

\bibitem[{{Chambers} {et~al.}(2016){Chambers}, {Magnier}, {Metcalfe},
  {Flewelling}, {Huber}, {Waters}, {Denneau}, {Draper}, {Farrow}, {Finkbeiner},
  {Holmberg}, {Koppenhoefer}, {Price}, {Rest}, {Saglia}, {Schlafly}, {Smartt},
  {Sweeney}, {Wainscoat}, {Burgett}, {Chastel}, {Grav}, {Heasley}, {Hodapp},
  {Jedicke}, {Kaiser}, {Kudritzki}, {Luppino}, {Lupton}, {Monet}, {Morgan},
  {Onaka}, {Shiao}, {Stubbs}, {Tonry}, {White}, {Ba{\~n}ados}, {Bell},
  {Bender}, {Bernard}, {Boegner}, {Boffi}, {Botticella}, {Calamida},
  {Casertano}, {Chen}, {Chen}, {Cole}, {Deacon}, {Frenk}, {Fitzsimmons},
  {Gezari}, {Gibbs}, {Goessl}, {Goggia}, {Gourgue}, {Goldman}, {Grant},
  {Grebel}, {Hambly}, {Hasinger}, {Heavens}, {Heckman}, {Henderson}, {Henning},
  {Holman}, {Hopp}, {Ip}, {Isani}, {Jackson}, {Keyes}, {Koekemoer}, {Kotak},
  {Le}, {Liska}, {Long}, {Lucey}, {Liu}, {Martin}, {Masci}, {McLean}, {Mindel},
  {Misra}, {Morganson}, {Murphy}, {Obaika}, {Narayan}, {Nieto-Santisteban},
  {Norberg}, {Peacock}, {Pier}, {Postman}, {Primak}, {Rae}, {Rai}, {Riess},
  {Riffeser}, {Rix}, {R{\"o}ser}, {Russel}, {Rutz}, {Schilbach}, {Schultz},
  {Scolnic}, {Strolger}, {Szalay}, {Seitz}, {Small}, {Smith}, {Soderblom},
  {Taylor}, {Thomson}, {Taylor}, {Thakar}, {Thiel}, {Thilker}, {Unger},
  {Urata}, {Valenti}, {Wagner}, {Walder}, {Walter}, {Watters}, {Werner},
  {Wood-Vasey}, \& {Wyse}}]{Chambers-2016}
{Chambers}, K.~C., {Magnier}, E.~A., {Metcalfe}, N., {et~al.} 2016, arXiv
  e-prints, arXiv:1612.05560.
\newblock \doarXiv{1612.05560}

\bibitem[{{Dey} {et~al.}(2019){Dey}, {Schlegel}, {Lang}, {Blum}, {Burleigh},
  {Fan}, {Findlay}, {Finkbeiner}, {Herrera}, {Juneau}, {Landriau}, {Levi},
  {McGreer}, {Meisner}, {Myers}, {Moustakas}, {Nugent}, {Patej}, {Schlafly},
  {Walker}, {Valdes}, {Weaver}, {Y{\`e}che}, {Zou}, {Zhou}, {Abareshi},
  {Abbott}, {Abolfathi}, {Aguilera}, {Alam}, {Allen}, {Alvarez}, {Annis},
  {Ansarinejad}, {Aubert}, {Beechert}, {Bell}, {BenZvi}, {Beutler}, {Bielby},
  {Bolton}, {Brice{\~n}o}, {Buckley-Geer}, {Butler}, {Calamida}, {Carlberg},
  {Carter}, {Casas}, {Castander}, {Choi}, {Comparat}, {Cukanovaite}, {Delubac},
  {DeVries}, {Dey}, {Dhungana}, {Dickinson}, {Ding}, {Donaldson}, {Duan},
  {Duckworth}, {Eftekharzadeh}, {Eisenstein}, {Etourneau}, {Fagrelius},
  {Farihi}, {Fitzpatrick}, {Font-Ribera}, {Fulmer}, {G{\"a}nsicke},
  {Gaztanaga}, {George}, {Gerdes}, {Gontcho}, {Gorgoni}, {Green}, {Guy},
  {Harmer}, {Hernandez}, {Honscheid}, {Huang}, {James}, {Jannuzi}, {Jiang},
  {Joyce}, {Karcher}, {Karkar}, {Kehoe}, {Kneib}, {Kueter-Young}, {Lan},
  {Lauer}, {Le Guillou}, {Le Van Suu}, {Lee}, {Lesser}, {Perreault Levasseur},
  {Li}, {Mann}, {Marshall}, {Mart{\'\i}nez-V{\'a}zquez}, {Martini}, {du Mas des
  Bourboux}, {McManus}, {Meier}, {M{\'e}nard}, {Metcalfe},
  {Mu{\~n}oz-Guti{\'e}rrez}, {Najita}, {Napier}, {Narayan}, {Newman}, {Nie},
  {Nord}, {Norman}, {Olsen}, {Paat}, {Palanque-Delabrouille}, {Peng},
  {Poppett}, {Poremba}, {Prakash}, {Rabinowitz}, {Raichoor}, {Rezaie},
  {Robertson}, {Roe}, {Ross}, {Ross}, {Rudnick}, {Safonova}, {Saha},
  {S{\'a}nchez}, {Savary}, {Schweiker}, {Scott}, {Seo}, {Shan}, {Silva},
  {Slepian}, {Soto}, {Sprayberry}, {Staten}, {Stillman}, {Stupak}, {Summers},
  {Sien Tie}, {Tirado}, {Vargas-Maga{\~n}a}, {Vivas}, {Wechsler}, {Williams},
  {Yang}, {Yang}, {Yapici}, {Zaritsky}, {Zenteno}, {Zhang}, {Zhang}, {Zhou}, \&
  {Zhou}}]{Dey-2019}
{Dey}, A., {Schlegel}, D.~J., {Lang}, D., {et~al.} 2019, \aj, 157, 168,
  \dodoi{10.3847/1538-3881/ab089d}

\bibitem[{{Drake} {et~al.}(2009){Drake}, {Djorgovski}, {Mahabal}, {Beshore},
  {Larson}, {Graham}, {Williams}, {Christensen}, {Catelan}, {Boattini},
  {Gibbs}, {Hill}, \& {Kowalski}}]{Drake-2009}
{Drake}, A.~J., {Djorgovski}, S.~G., {Mahabal}, A., {et~al.} 2009, \apj, 696,
  870, \dodoi{10.1088/0004-637X/696/1/870}

\bibitem[{{Huertas-Company} {et~al.}(2011){Huertas-Company}, {Aguerri},
  {Bernardi}, {Mei}, \& {S{\'a}nchez Almeida}}]{Huertas-2011}
{Huertas-Company}, M., {Aguerri}, J.~A.~L., {Bernardi}, M., {Mei}, S., \&
  {S{\'a}nchez Almeida}, J. 2011, \aap, 525, A157,
  \dodoi{10.1051/0004-6361/201015735}

\bibitem[{{Li} {et~al.}(2011){Li}, {Chornock}, {Leaman}, {Filippenko},
  {Poznanski}, {Wang}, {Ganeshalingam}, \& {Mannucci}}]{Li-2011}
{Li}, W., {Chornock}, R., {Leaman}, J., {et~al.} 2011, \mnras, 412, 1473,
  \dodoi{10.1111/j.1365-2966.2011.18162.x}

\bibitem[{{Liske} {et~al.}(2015){Liske}, {Baldry}, {Driver}, {Tuffs},
  {Alpaslan}, {Andrae}, {Brough}, {Cluver}, {Grootes}, {Gunawardhana},
  {Kelvin}, {Loveday}, {Robotham}, {Taylor}, {Bamford}, {Bland-Hawthorn},
  {Brown}, {Drinkwater}, {Hopkins}, {Meyer}, {Norberg}, {Peacock}, {Agius},
  {Andrews}, {Bauer}, {Ching}, {Colless}, {Conselice}, {Croom}, {Davies}, {De
  Propris}, {Dunne}, {Eardley}, {Ellis}, {Foster}, {Frenk}, {H{\"a}u{\ss}ler},
  {Holwerda}, {Howlett}, {Ibarra}, {Jarvis}, {Jones}, {Kafle}, {Lacey},
  {Lange}, {Lara-L{\'o}pez}, {L{\'o}pez-S{\'a}nchez}, {Maddox}, {Madore},
  {McNaught-Roberts}, {Moffett}, {Nichol}, {Owers}, {Palamara}, {Penny},
  {Phillipps}, {Pimbblet}, {Popescu}, {Prescott}, {Proctor}, {Sadler},
  {Sansom}, {Seibert}, {Sharp}, {Sutherland}, {V{\'a}zquez-Mata}, {van Kampen},
  {Wilkins}, {Williams}, \& {Wright}}]{Liske-2015}
{Liske}, J., {Baldry}, I.~K., {Driver}, S.~P., {et~al.} 2015, \mnras, 452,
  2087, \dodoi{10.1093/mnras/stv1436}

\bibitem[{{Matheson} {et~al.}(2021){Matheson}, {Stubens}, {Wolf}, {Lee},
  {Narayan}, {Saha}, {Scott}, {Soraisam}, {Bolton}, {Hauger}, {Silva},
  {Kececioglu}, {Scheidegger}, {Snodgrass}, {Aleo}, {Evans-Jacquez}, {Singh},
  {Wang}, {Yang}, \& {Zhao}}]{Matheson-2021}
{Matheson}, T., {Stubens}, C., {Wolf}, N., {et~al.} 2021, \aj, 161, 107,
  \dodoi{10.3847/1538-3881/abd703}

\bibitem[{{Mauerhan} {et~al.}(2013){Mauerhan}, {Smith}, {Filippenko},
  {Blanchard}, {Blanchard}, {Casper}, {Cenko}, {Clubb}, {Cohen}, {Fuller},
  {Li}, \& {Silverman}}]{Mauerhan-2013}
{Mauerhan}, J.~C., {Smith}, N., {Filippenko}, A.~V., {et~al.} 2013, \mnras,
  430, 1801, \dodoi{10.1093/mnras/stt009}

\bibitem[{{Ofek} {et~al.}(2013){Ofek}, {Sullivan}, {Cenko}, {Kasliwal},
  {Gal-Yam}, {Kulkarni}, {Arcavi}, {Bildsten}, {Bloom}, {Horesh}, {Howell},
  {Filippenko}, {Laher}, {Murray}, {Nakar}, {Nugent}, {Silverman}, {Shaviv},
  {Surace}, \& {Yaron}}]{Ofek-2013}
{Ofek}, E.~O., {Sullivan}, M., {Cenko}, S.~B., {et~al.} 2013, \nat, 494, 65,
  \dodoi{10.1038/nature11877}

\bibitem[{{Phillips} {et~al.}(1992){Phillips}, {Wells}, {Suntzeff}, {Hamuy},
  {Leibundgut}, {Kirshner}, \& {Foltz}}]{Phillips-1992}
{Phillips}, M.~M., {Wells}, L.~A., {Suntzeff}, N.~B., {et~al.} 1992, \aj, 103,
  1632, \dodoi{10.1086/116177}

\bibitem[{{Planck Collaboration} {et~al.}(2014){Planck Collaboration}, {Ade},
  {Aghanim}, {Armitage-Caplan}, {Arnaud}, {Ashdown}, {Atrio-Barandela},
  {Aumont}, {Baccigalupi}, {Banday}, {Barreiro}, {Bartlett}, {Battaner},
  {Benabed}, {Beno{\^\i}t}, {Benoit-L{\'e}vy}, {Bernard}, {Bersanelli},
  {Bielewicz}, {Bobin}, {Bock}, {Bonaldi}, {Bond}, {Borrill}, {Bouchet},
  {Bridges}, {Bucher}, {Burigana}, {Butler}, {Calabrese}, {Cappellini},
  {Cardoso}, {Catalano}, {Challinor}, {Chamballu}, {Chary}, {Chen}, {Chiang},
  {Chiang}, {Christensen}, {Church}, {Clements}, {Colombi}, {Colombo},
  {Couchot}, {Coulais}, {Crill}, {Curto}, {Cuttaia}, {Danese}, {Davies},
  {Davis}, {de Bernardis}, {de Rosa}, {de Zotti}, {Delabrouille}, {Delouis},
  {D{\'e}sert}, {Dickinson}, {Diego}, {Dolag}, {Dole}, {Donzelli}, {Dor{\'e}},
  {Douspis}, {Dunkley}, {Dupac}, {Efstathiou}, {Elsner}, {En{\ss}lin},
  {Eriksen}, {Finelli}, {Forni}, {Frailis}, {Fraisse}, {Franceschi}, {Gaier},
  {Galeotta}, {Galli}, {Ganga}, {Giard}, {Giardino}, {Giraud-H{\'e}raud},
  {Gjerl{\o}w}, {Gonz{\'a}lez-Nuevo}, {G{\'o}rski}, {Gratton}, {Gregorio},
  {Gruppuso}, {Gudmundsson}, {Haissinski}, {Hamann}, {Hansen}, {Hanson},
  {Harrison}, {Henrot-Versill{\'e}}, {Hern{\'a}ndez-Monteagudo}, {Herranz},
  {Hildebrandt}, {Hivon}, {Hobson}, {Holmes}, {Hornstrup}, {Hou}, {Hovest},
  {Huffenberger}, {Jaffe}, {Jaffe}, {Jewell}, {Jones}, {Juvela},
  {Keih{\"a}nen}, {Keskitalo}, {Kisner}, {Kneissl}, {Knoche}, {Knox}, {Kunz},
  {Kurki-Suonio}, {Lagache}, {L{\"a}hteenm{\"a}ki}, {Lamarre}, {Lasenby},
  {Lattanzi}, {Laureijs}, {Lawrence}, {Leach}, {Leahy}, {Leonardi},
  {Le{\'o}n-Tavares}, {Lesgourgues}, {Lewis}, {Liguori}, {Lilje},
  {Linden-V{\o}rnle}, {L{\'o}pez-Caniego}, {Lubin}, {Mac{\'\i}as-P{\'e}rez},
  {Maffei}, {Maino}, {Mandolesi}, {Maris}, {Marshall}, {Martin},
  {Mart{\'\i}nez-Gonz{\'a}lez}, {Masi}, {Massardi}, {Matarrese}, {Matthai},
  {Mazzotta}, {Meinhold}, {Melchiorri}, {Melin}, {Mendes}, {Menegoni},
  {Mennella}, {Migliaccio}, {Millea}, {Mitra}, {Miville-Desch{\^e}nes},
  {Moneti}, {Montier}, {Morgante}, {Mortlock}, {Moss}, {Munshi}, {Murphy},
  {Naselsky}, {Nati}, {Natoli}, {Netterfield}, {N{\o}rgaard-Nielsen},
  {Noviello}, {Novikov}, {Novikov}, {O'Dwyer}, {Osborne}, {Oxborrow}, {Paci},
  {Pagano}, {Pajot}, {Paladini}, {Paoletti}, {Partridge}, {Pasian},
  {Patanchon}, {Pearson}, {Pearson}, {Peiris}, {Perdereau}, {Perotto},
  {Perrotta}, {Pettorino}, {Piacentini}, {Piat}, {Pierpaoli}, {Pietrobon},
  {Plaszczynski}, {Platania}, {Pointecouteau}, {Polenta}, {Ponthieu}, {Popa},
  {Poutanen}, {Pratt}, {Pr{\'e}zeau}, {Prunet}, {Puget}, {Rachen}, {Reach},
  {Rebolo}, {Reinecke}, {Remazeilles}, {Renault}, {Ricciardi}, {Riller},
  {Ristorcelli}, {Rocha}, {Rosset}, {Roudier}, {Rowan-Robinson},
  {Rubi{\~n}o-Mart{\'\i}n}, {Rusholme}, {Sandri}, {Santos}, {Savelainen},
  {Savini}, {Scott}, {Seiffert}, {Shellard}, {Spencer}, {Starck}, {Stolyarov},
  {Stompor}, {Sudiwala}, {Sunyaev}, {Sureau}, {Sutton}, {Suur-Uski}, {Sygnet},
  {Tauber}, {Tavagnacco}, {Terenzi}, {Toffolatti}, {Tomasi}, {Tristram},
  {Tucci}, {Tuovinen}, {T{\"u}rler}, {Umana}, {Valenziano}, {Valiviita}, {Van
  Tent}, {Vielva}, {Villa}, {Vittorio}, {Wade}, {Wandelt}, {Wehus}, {White},
  {White}, {Wilkinson}, {Yvon}, {Zacchei}, \& {Zonca}}]{planck-13}
{Planck Collaboration}, {Ade}, P.~A.~R., {Aghanim}, N., {et~al.} 2014, \aap,
  571, A16, \dodoi{10.1051/0004-6361/201321591}

\bibitem[{{Schlafly} \& {Finkbeiner}(2011)}]{Schlafly-2011}
{Schlafly}, E.~F., \& {Finkbeiner}, D.~P. 2011, \apj, 737, 103,
  \dodoi{10.1088/0004-637X/737/2/103}

\bibitem[{{Scolnic} {et~al.}(2020){Scolnic}, {Smith}, {Massiah}, {Wiseman},
  {Brout}, {Kessler}, {Davis}, {Foley}, {Galbany}, {Hinton}, {Hounsell},
  {Kelsey}, {Lidman}, {Macaulay}, {Morgan}, {Nichol}, {M{\"o}ller}, {Popovic},
  {Sako}, {Sullivan}, {Thomas}, {Tucker}, {Abbott}, {Aguena}, {Allam}, {Annis},
  {Avila}, {Bechtol}, {Bertin}, {Brooks}, {Burke}, {Rosell}, {Carollo}, {Kind},
  {Carretero}, {Costanzi}, {da Costa}, {De Vicente}, {Desai}, {Diehl}, {Doel},
  {Drlica-Wagner}, {Eckert}, {Eifler}, {Everett}, {Flaugher}, {Fosalba},
  {Frieman}, {Garc{\'\i}a-Bellido}, {Gaztanaga}, {Gerdes}, {Glazebrook},
  {Gruen}, {Gruendl}, {Gschwend}, {Gutierrez}, {Hartley}, {Hollowood},
  {Honscheid}, {James}, {Kuehn}, {Kuropatkin}, {Lewis}, {Li}, {Lima}, {Maia},
  {Marshall}, {Menanteau}, {Miquel}, {Palmese}, {Paz-Chinch{\'o}n}, {Plazas},
  {Pursiainen}, {Sanchez}, {Scarpine}, {Schubnell}, {Serrano},
  {Sevilla-Noarbe}, {Sommer}, {Suchyta}, {Swanson}, {Tarle}, {Varga}, {Walker},
  {Wilkinson}, \& {DES Collaboration}}]{Scolnic-2020}
{Scolnic}, D., {Smith}, M., {Massiah}, A., {et~al.} 2020, \apjl, 896, L13,
  \dodoi{10.3847/2041-8213/ab8735}

\bibitem[{{Soraisam} \& {Gilfanov}(2015)}]{Soraisam-2015}
{Soraisam}, M.~D., \& {Gilfanov}, M. 2015, \aap, 583, A140,
  \dodoi{10.1051/0004-6361/201424118}

\bibitem[{{Tonry} {et~al.}(2018){Tonry}, {Denneau}, {Heinze}, {Stalder},
  {Smith}, {Smartt}, {Stubbs}, {Weiland}, \& {Rest}}]{Tonry-2018}
{Tonry}, J.~L., {Denneau}, L., {Heinze}, A.~N., {et~al.} 2018, \pasp, 130,
  064505, \dodoi{10.1088/1538-3873/aabadf}

\bibitem[{{Wambsganss}(1998)}]{Wambsganss-1998}
{Wambsganss}, J. 1998, Living Reviews in Relativity, 1, 12,
  \dodoi{10.12942/lrr-1998-12}

\end{thebibliography}
\bibliographystyle{aasjournal}

\end{document}